\documentclass[runningheads]{llncs}
\usepackage[T1]{fontenc}
\usepackage[utf8]{inputenc}
\usepackage{graphicx}
\usepackage{subfig}
\usepackage{fixme}
\fxsetup{status=draft}

\usepackage{hyperref}

\begin{document}
%
%
%
\author{Diogo Lemos \inst{q}\orcidID{0000-0000-0000-0000} 
\and Ademar Aguiar\inst{1}\orcidID{0000-0002-4046-4729}
\and Neil B. Harrison\inst{2}\orcidID{0009-0007-4620-8818} 
 }
%
%
\institute{INESC TEC, Faculdade de Engenharia da Universidade do Porto, Portugal \\ 
\email{up202003484@fe.up.pt, ademar.aguiar@fe.up.pt}
\and Utah Valley University, Orem Utah 84058, USA \\
\email{neil.harrison@uvu.edu} \\
}

\title{Requirements for Active Assistance of \\Natural Questions in Software Architecture}
\titlerunning{Active Assistance of \\Natural Questions in Software Architecture}

\maketitle

\begin{abstract}

Natural questions are crucial to shaping key architectural decisions and preserving architectural knowledge.
They arise organically during the architectural design process, often resulting from the existing architectural experience of the designer and the distinctive characteristics of the system being designed.
However, natural questions are often mismanaged or ignored, which can lead to architectural drift, knowledge loss, inefficient resource use, or poor understandability of the system's architecture.
We aim to better understand the lifecycle of natural questions, its key requirements, challenges and difficulties, and then to envision an assisted environment to properly support it. 
The environment should be adaptable and responsive to real-world constraints and uncertainties by seamlessly integrating knowledge management tools and artificial intelligence techniques into software development workflows.
Based on existing literature, a requirements workshop, and three design iterations, we proposed a lifecycle for natural questions and elicited essential functional and non-functional requirements for such an environment.
At last, the results of a survey conducted with experts helped to analyze and validate the elicited requirements and proposed features for the environment to enhance collaboration, decision-making, and the preservation of architectural knowledge more effectively than conventional methods.
\end{abstract}

\keywords{Software Architecture \and Requirements \and Quality Attributes}

\newpage

\section{Introduction}
During the software architecture design process, questions arise spontaneously from architects, developers, researchers, and stakeholders as they strive for a better understanding of requirements, long-term consequences, and tradeoffs.
Each question frequently contains additional information beyond its response, potentially capturing the context as well. 
Altogether, these questions can contribute to a more comprehensive understanding of the overall system. We designate such spontaneous questions as 'natural questions'.

\subsubsection{Importance.}
The importance of natural questions is well understood. Since software architecture is inherently decision-driven, these questions often significantly influence key decisions and critical choices shaping system design, implementation and evolution.
Unfortunately, despite their importance, naturally occurring questions are frequently overlooked, poorly managed, informally documented, and even lost, when not properly supported.

\subsubsection{Lifecycle.}
Questions are dynamic and arise in various forms that must be systematically collected and maintained. These can take shape in written conversations, discussed via speech and shared through diagrams and images, making the storage and organization of knowledge demanding and laborious.
Gathering and maintaining as much as possible of every phase each question goes through is extremely important if we want to preserve high-fidelity, easily accessible architectural knowledge.

\subsubsection{Management.}
Ineffective question management can be severe and lead to costly errors.
Without a tailored and structured solution to capture and preserve architectural questions, systems may be more susceptible to numerous difficulties, including knowledge loss, misaligned decisions, inconsistent and often outdated records, and communication bottlenecks \cite{harrisonaguiar2024nature}.
These challenges may manifest themselves throughout the system’s development, affecting its maintenance, evolution, and scalability. Consequently, they can result in a substantial amount of resources being wasted\cite{tang}.

\subsubsection{Tools.}
However, their tracking is loosely managed due to the lack of accessible, easy-to-use solutions. 
Considering the dynamic and spontaneous nature of questions, either ad hoc documentation methods or semi-structured but informal ones (such as post-its, meeting notes, emails, diagrams, and specifications) may not be the most efficient and effective to address the challenges of systematically supporting the evolution of the architectural knowledge captured by questions~\cite{Gilson}.
In a previous study \cite{harrisonaguiar2024nature}, we found that the tracking and organisation of natural questions is multiple dimensional with the often use of ad hoc methods, with some mentions of platforms, such as \textit{Jira} and \textit{Trello}, and informal methods, for example, to-do lists. 
While not specialised for architectural questions, these methods are also the probable cause of the lack of preservation of architectural knowledge.
To address these issues, the study recommends the development of a specialised set of tools that can support the eliciting, organising and tracking of natural questions in software architectural processes.

\subsubsection{Active Assisted Lifecycle.}
The aim of this work is twofold: to understand in detail the lifecycle of natural questions and its requirements and then later to explore novel ways and tools to properly assist it.

In this paper, we present the results of a study conducted using a mix of research methods to identify the key requirements to adequately support the lifecycle of the natural questions that arise during software architecture design. The resulting requirements will drive the design of a preliminary environment, which we designate here as an active assisted environment. 

We began the work by mining the lifecycle of questions from existing literature (roles, phases, and activities). 
Subsequently, we gathered the key functional and non-functional requirements to support the lifecycle through a workshop with 50 students from a software architecture course at the University of Porto. 
We then conducted an online survey to analyse the requirements, targeting architecture experts, for which we received approximately 18 responses. 
Based on the findings, we started drafting an initial conceptual architecture and user interfaces for the environment, which is yet to be refined, prototyped, and validated in the future.


\section{Research Methodology}
A literature review was previously conducted by the authors to comprehensively understand the past and current landscape of software architecture questions, knowledge management, decision traceability, and solutions that could support our research objectives. 

To investigate the potential of natural questions for preserving and comprehending knowledge pertaining to software architecture, the primary research questions addressed in this work were as follows:

\begin{description}
    \item[RQ1.] What is the \textit{typical lifecycle} of natural questions that emerge during the software architecture process?
    \item[RQ2.] What are the \textit{fundamental requirements} for systematically supporting the lifecycle of natural questions?
\end{description}




To answer the research questions, we followed a qualitative research approach using two complementary empirical methods: (i) a requirements workshop with software architecture students, mainly for elicitation of functional and non-functional requirements, for which we suggested the use of natural questions; and (ii) a survey with experienced software architects, mainly for requirements analysis and prioritization.

\begin{figure}[h]
    \centering
    \includegraphics[width=1\textwidth]{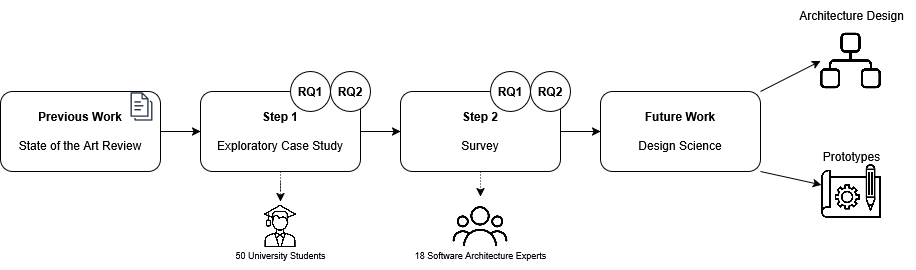} %
    \caption{Research Methodology}
    \label{fig:researchmethod}
\end{figure}

\section{Related Work}



Drawing from knowledge management theories, such as the Ba concept \cite{conceptba,conceptba-aguiar}, we recognize that architectural questions are not static entities but dynamic interactions that evolve through collaborative spaces. 
This perspective emphasizes the importance of supporting the continuous transformation of tacit and explicit knowledge, particularly in capturing and understanding architectural questions.

This section delves into the existing solutions, studies, contributions and limitations regarding knowledge gathering, tracking and preservation in software architecture.

\subsection{Architectural Knowledge Preservation} 
As software development and engineering evolve in levels of complexity, one critical concern for organisations is the preservation of architectural knowledge~(AK), given its crucial role in the successful development and maintenance of every software system. Architectural knowledge encompasses design decisions \cite{Jansen2005}, architectural solutions, patterns and principles that shape systems' overall structure and evolution. 


Architectural Decision Records~(ADRs) are considered critical in the documentation and monitoring of key decisions in software architecture, ultimately promoting collaboration and understanding~\cite{Vaidhyanathan}. 
While ADRs provide a structured format for recording finalised architectural choices, they cannot capture the complex decision-making process, which often involves questions, assumptions, debates, and evolving insights.


This limited information before a decision results in limited context when revisiting past architectural choices.
The tools, such as Trello and Jira, attempt to fulfil knowledge management.
However, since they lack specialisation in architectural knowledge, they fail to track the dynamic evolution of architectural questions, resulting in scattered, informal and difficult-to-retrieve knowledge, reinforcing the need for a dedicated system that integrates question tracking as crucial than simply dealing with the identified problems that come with it \cite{Dhar}.

\subsection{AI for Architectural Knowledge}

In software architecture, the preservation of architectural knowledge and the utilization of artificial intelligence (AI) as a means to aid in its maintenance have been extensively explored. 

Recent advancements in AI have introduced new ways to manage architectural knowledge and transformed how it is retrieved and documented\cite{Drozdzewski}. Works such as Borrego et al. \cite{Borrego} propose an NLP-based tagging system that would classify architectural knowledge in agile teams, enhancing the capture and organisation of knowledge.

As suggested in the work by Dhar et al.~\cite{Vaidhyanathan}, Large Language Models trained on architectural patterns would be able to assist in the creation of ADRs but fail to capture design decisions at a human level comprehensively.
Natural questions and their discussions involve contextual ambiguities, conflicting viewpoints and ever-evolving considerations, requiring a more specialised system that, through varied tools, would support question management, tracking and resolution over time.

From the reviewed works, several key limitations emerge. 
Despite AI's proven utility in software engineering applications like code generation \cite{Kokol}, existing tools fall short in dynamically tracking, classifying, and prioritising architectural knowledge. 
The fragmentation of knowledge storage also appears challenging due to the informal and disconnected methods that are used to communicate and maintain crucial knowledge, making it difficult to trace the evolution of architectural questions and their long-term implications.


\subsection{Open issues}

Despite significantly contributing to structuring architectural knowledge, these contributions fail to systematically support the dynamic lifecycle of natural questions, leaving a gap when capturing, organising and preserving these crucial questions in software systems. 

Given the highlighted gaps, specialised tools that systematically support the lifecycle of natural questions can offset many challenges present in current methods. 

Rather than focusing on decisions, it should integrate natural question tracking from inception to resolution by leveraging AI to categorise, prioritise and retrieve architectural questions.

Such tools should focus on delivering an efficient knowledge-sharing solution while ensuring traceability between questions and decisions by integrating seamlessly into collaborative workflows.

\section{Lifecycle of Natural Questions}

Understanding the dynamic lifecycle of natural questions is vital for their preservation and management. Using an action research method, we explored how these questions arise, modify and influence decision-making throughout the software development process. Modelling the lifecycle ensures that the proposed tool supports the entire life span of an architectural question, from its inception to its archival, ensuring its eventual long-term impact. 

To better understand the problem, we depicted in an activity diagram the roles and activities surrounding a question's lifecycle. 
For a generalised software system, we regard four roles as transversal from any system since they follow the dynamic nature of the questions' lifecycle. 
These roles are Question Owner, Product Owner, Developer/Researcher and Decision Maker.

\begin{description}
    \item[Questions Owner.] Formulates the question and introduces it into the workflow, discussing it with other members to ensure it is clearly defined while also managing the question backlog and archival process.
    \item[Product Owner.] Assess the priority of the question, aligning it with system objectives, being a part of the discussion and reflection phase. When a question re-emerges, it is key to start the process again. This role fits well with the Product Owner from Scrum \cite{sutherland_scrum_2019}.
    \item[Developers/Researchers.] Conduct in-depth analyses and delve deeper to provide insights and solutions for discussion with other members.
    \item[Decision Makers.] Evaluate all parameters and determine whether the question is resolved, deferred, or answered through assumptions.
\end{description}

The identification of these specific roles represents a nuanced approach to understanding architectural question management. 
While drawing from existing software systems management literature, these roles emerged through a systematic analytical process of deconstructing software development and architectural decision-making practices. 
This framework recognises that architectural questions involve complex interactions that cannot be captured by traditional linear models.

It is important to note that while the proposed lifecycle presents a typical path of question management, it is not prescriptive. In practice, question resolution may occasionally deviate from this model, with potential different role interactions or unconventional routes that bypass standard communication channels. The framework recognises this flexibility, acknowledging that real-world software development processes often require adaptive communication strategies.

\begin{figure}[h]
    \centering
    \includegraphics[width=1.0\textwidth]{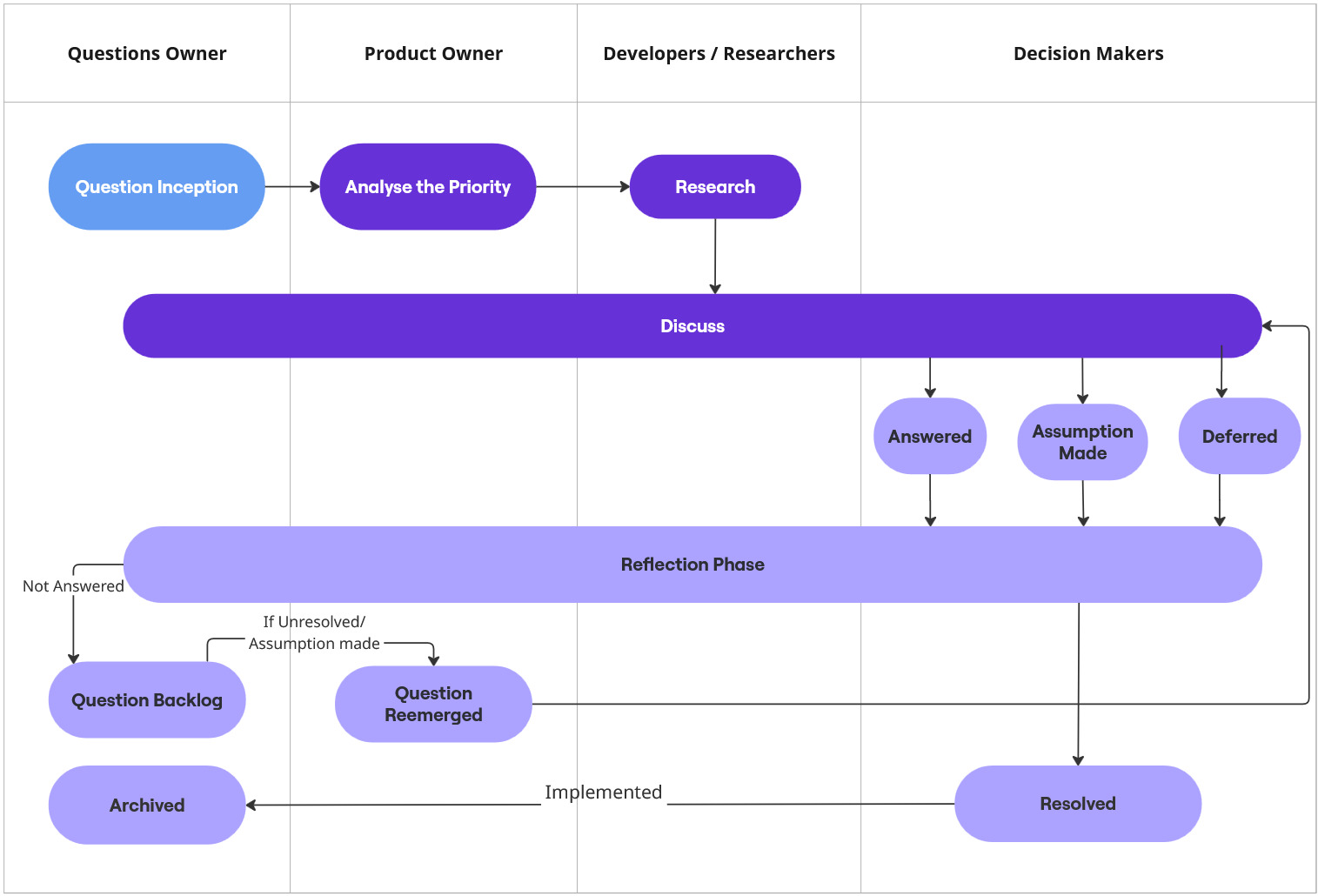} %
    \caption{Lifecycle of a Question}
    \label{fig:question_lifecycle}
\end{figure}

The phases highlighted in Figure~\ref{fig:question_lifecycle} consider the ongoing discussions for each question, often evolving multiple roles and the three possible outcomes of resolving a question. 
A question may be answered and resolved, it may lack a concrete and final answer at an early stage but have an answer assumed to keep the system design and development moving forward, or it may remain unanswered at a given stage due to uncertainties or constraints.

\section{Requirements Workshop} 

A workshop involving 50 university students from a master’s course on Software Systems Architecture at the University of Porto \cite{softarch2025} was conducted to support the process of eliciting, validating and refining the requirements for supporting tools. This workshop allowed us to better understand how different natural questions can be perceived and managed and gather feedback on potential requirements and respective tools' features. 
The key activities performed in this workshop were:

\begin{itemize}
    \item Acting as software architects, the students wrote down questions that arose while designing an architecture.
    \item The students grouped related questions using sticky notes and visualised connections through spiderweb diagrams.
    \item Peer and instructor feedback was key to improving clarity and usability when presenting their findings.
\end{itemize}

\begin{figure}[!thbp]
  \centering
  {\includegraphics[width=1.0\textwidth]{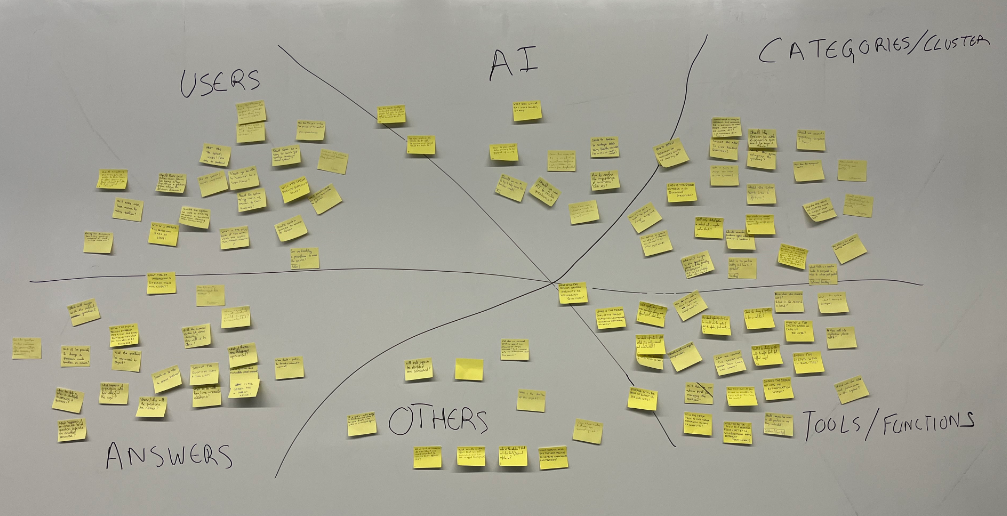}\label{fig:f1}}
  \caption{Example of one of the spiderwebs.}
  \label{fig:spiderweb}
\end{figure}

This collaborative effort structured the questions into six main categories: Users, AI, Categories, Tools/Functions, Answers, and Others. 
Each category captured distinct concerns, such as user needs, question classification, role attribution, tool capabilities and AI's role (see Figure~\ref{fig:spiderweb}).

Several crucial questions were raised, and with them, newfound knowledge was achieved, once again highlighting the importance of questions in shaping software architecture decisions. 
The structured discussions uncovered critical gaps in current question management practices and reinforced the necessity of a systematic approach to capturing and tracking architectural questions. 

The insights gained from this workshop played a pivotal role in refining the tools' requirements, ensuring they effectively support decision-making, knowledge preservation, and the overall evolution of software architecture.





\section{Requirements for an Assisted Environment}

The objective of the envisioned supporting tools is to overcome the divide between unstructured question management and structured architectural knowledge preservation, guaranteeing that vital decisions are well-documented and readily available for future usage. 
By addressing these challenges, our work contributes to establishing a structured and actively assisted approach to managing architectural questions, ensuring informed decision-making, reducing unnecessary rework, and enhancing the preservation and accessibility of architectural knowledge.

A requirement-driven approach was adopted to envision the tools and analyse the core functionalities necessary for such an environment, including question prioritisation, retrieval, decision traceability and key insights extraction from several sources. 

The requirements were identified by analysing the needs of the key roles described previously, performing a gap analysis of existing solutions, and gathering insights from a plethora of stakeholders and university students.


The main gaps encountered were a lack of structured question tracking, fragmented documentation procedures and limited context preservation. Natural questions are qualitatively different than action items, requiring different ways to be managed and tracked: action items are tasks to complete something, usually known and natural questions are by nature unknowns and, therefore, often involve risks.
Even if action items are to find something out, they are just one type of action item. Because of the risk, the inherent unknowns, and the specialized lifecycle, they need special handling. 

Based on these insights, we specified functional and non-functional requirements to address these limitations and explore the potential of structured knowledge management and seamless collaboration in software architecture workflows.


\subsection{Functional Requirements}
The functional requirements (FRs) were organized into four categories and prioritized to ensure the full capacity of the intended tools.

\subsubsection{Questions and Knowledge Retrieval}

\begin{description}
    \item[FR1.] The system must allow users to register questions via text and voice input to ensure that knowledge is captured in various formats, ensuring easiness of use.
    \item[FR2.] The system should recognise images and diagrams, extracting relevant insights for question analysis and discussion.
    \item[FR3.] The system should transcribe audio recordings into structured text, automating transcription and filtering out irrelevant content to improve efficiency.
    \item[FR4.] The tool must provide searching, allowing for the retrieval of past questions through keywords, priority and issue.
\end{description}

\subsubsection{Classification, Prioritisation and Tracking}

\begin{description}
    \item[FR5.] The system must classify and store questions based on predefined categories and priorities to facilitate retrieval.
    \item[FR6.] The system must maintain a question backlog consisting of active, resolved, assumed and archived questions.
    \item[FR7.] The system must track decision-making processes, associating questions with choices and contributors.
    \item[FR8.] The system must allow Product owners to prioritise questions based on urgency and impact to improve decision-making and resource allocation.
\end{description}

\subsubsection{AI-driven Insights}

\begin{description}
    \item[FR9.] The system must detect and highlight similar or ambiguous questions to reduce redundancy.
    \item[FR10.] The system must provide AI-generated recommendations for 
    addressing architectural concerns and common pitfalls.
\end{description}

\subsubsection{Collaboration and Security}
\begin{description}
    \item[FR11.] The system should support multi-platform access to ensure availability.
    \item[FR12.] The system must notify users when a question is updated, commented or resolved to enhance collaboration.
    \item[FR13.] The system must implement role-based access control to restrict information visibility based on user roles.
\end{description}

\subsection{Non-Functional Requirements}

To ensure high security, performance and usability, in addition to functional requirements, the system must meet these non-functional requirements (NFRs).

\begin{description}
    \item[NFR1] The system must quickly process user queries and responses to ensure real-time usability.
    \item[NFR2] The user interface must be intuitive for any role, requiring a small number of interactions for each action regardless of the role.
    \item[NFR3] The system must be accessible anytime and anywhere, supporting the users and allowing spur-of-the-moment questions to be raised and addressed.
    \item[NFR4] All questions, decisions, and interactions must be secure, ensuring data integrity.
    \item[NFR5] The system should implement multi-level security, granting access based on user roles.
\end{description}

\section{Survey with Experienced Architects}

Feedback was gathered from experienced software architects, developers and university students at different stages of the design process to analyse, validate and ensure the relevance and feasibility of these requirements. 
Initially, through the workshop, key insights were highlighted, heightening the understanding of the improvements and requirements necessary to create a relevant and complete environment for natural questions. The most crucial insights suggest that:

\begin{itemize}
    \item AI tools and AI-based classification and prioritisation are key differentiators from existing tools, such as those for action items.
    \item Multi-platform support that ensures accessibility is crucial for the adoption across teams and members.
    \item Questions were raised on how much the system should intervene during discussion/debate phases.
\end{itemize}

While the workshop provided valuable insights from students, further validation from experienced software architecture experts was essential to evaluate the feasibility and industry relevance of the proposed requirements. 
To achieve this, we conducted a survey targeting professionals in the field, aiming to refine and validate our approach with expert insights and collect both quantitative and qualitative data on the lifecycle of architectural questions, the drawbacks associated with their management and the potential benefits of a structured tool for systematic tracking.

\subsubsection{Demographics.}
The survey was aimed at experienced software architects with varying levels of experience and involvement in projects of various scales. 
A total of 18 responses were collected, with 16 respondents having at least 10 years of experience in software development and 11 having more than 20 years of experience. 
As software architects, only 6 had less than five years of experience, and 10 of the inquired had at least 10 years of experience in this role. The participants worked on projects of various sizes, ranging from small-scale applications to large enterprise systems. The distribution of project sizes highlights a preponderance of medium-sized projects with the majority of answers being shared between medium and small scale projects.

\subsection{Challenges in Managing Architectural Questions}
Participants were asked if they encountered challenges in managing architectural questions systematically, which resulted in 16 participants voting yes. Participants were then asked to identify the most significant challenges when systematically managing architectural questions. The results were distributed across all options and the top three challenges reported were:

\begin{description}
    \item[Fragmented documentation practices]suggest that existing documentation methods are inconsistent and lead to scattered information across different tools, documents and repositories. This fragmentation is a major challenge when tracking and referencing architectural questions, discussions and decisions.
    
    \item[Difficulty retrieving past decisions]indicates that important architectural knowledge is often hard to search for due to a lack of proper management of questions that would render them available at all times.
    
    \item[Knowledge loss over time]~emphasizes that architectural insights and reasoning behind decisions tend to disappear when not properly recorded, especially in long-term projects or projects with rotating roles between members.

\end{description}

\begin{figure}[h]
    \centering
    \includegraphics[width=1.0\textwidth]{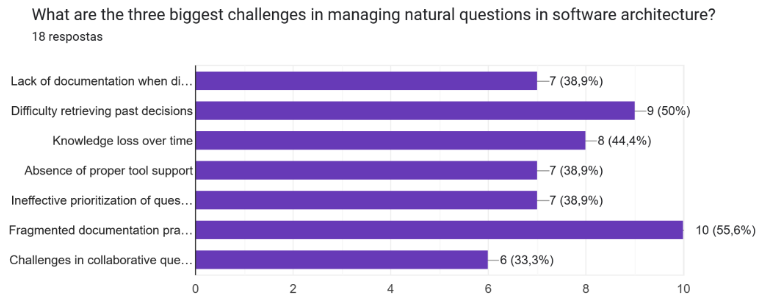} %
    \label{fig:challenges}
    \caption{Biggest challenges identified.}
\end{figure}

These findings reinforce the critical need for a structured approach to managing architectural questions, ensuring that key discussions and decisions remain accessible, well-documented, and preserved throughout the software development lifecycle.

\subsection{Roles in the Architectural Question Lifecycle}

Participants were asked to evaluate the proposed roles involved in the architectural questions lifecycle presented in the form of the activity diagram. The respondents were divided between full agreement (10) and partial agreement (7) with the roles displayed, with only one person not agreeing with them.

Key insights from this section include the partial agreement with the roles presented indicating general alignment with the role structure while also raising interesting suggestions and refinements regarding the role allocation. Some participants raised concerns about the role of the Product Owner, questioning their participation in revisiting questions or if another role should handle this responsibility. These concerns suggest that while the roles are agreed upon at a larger scale, there may be a need for further clarification regarding responsibilities.

\subsection{Activities in the Architectural Question Lifecycle}

Similar to the previous questions, we inquired the software architects on their opinion relating to the activities displayed in the activity diagram. The majority (17) found this classification useful and mostly accurate, with 9 agreeing completely with what was presented.

The classification was deemed effective in capturing how questions evolve in practice, acknowledging the iterative nature of questions. The activity diagram resonated well with participants who valued its clarity in mapping the status of questions. However, some participants suggested the addition of impact assessment activities when a question re-emerges and also a validation stage. These findings validated our classification since it was generally well-received, but it could benefit from additional granularity or refinements as stated previously. This could offer a more nuanced view of how questions evolve over time and ensure better tracking in the lifecycle. 

\subsection{Necessity for Tracking Architectural Decisions}

When inquired to rate the importance of systematically tracking architectural questions, every participant gave at least a grade of 3 on a scale of 1 to 5, with 8 people rating it a 5 out of 5, suggesting that there is a broad agreement on how important the tracking of these questions is. Following this question, the architects were then asked if there was a need for a structured tool for managing natural questions in software architecture, which was answered yes by every participant barring one (17).

\subsection{Ranking of Necessary Features}

After observing the resounding majority of participants agreeing to the necessity of a structured tool to manage natural questions in software architecture, we inquired about what features the software architects regarded as the most important ones. The three most requested features, ranked in order of priority, were:

\begin{description}
    \item[Question backlog management to track answered and unresolved ones.] Ensuring that all architectural questions are systematically recorded, monitored and revisited when necessary.
    \item[Architecture diagram recognition for visual knowledge extraction.] Enabling the interpretation and extraction of insights from architectural diagrams, providing a more integrated approach to managing architectural knowledge in the environment.
    \item[Conversational AI assistant to answer questions in natural language.]Fostering quick and intuitive access to historical architectural knowledge through an AI-powered assistant capable of interpreting and answering natural language questions. 
\end{description}

These insights validate and display the requirements and features that architects prioritize for managing architectural questions, discussions and decisions. With this survey, a number of requirements elicited before were validated and their prioritization is now supported.

\subsection{Improved Lifecycle of Natural Questions}

Following the insights shared by software architecture experts, we carefully analysed all suggestions regarding the question lifecycle activity diagram. While the expert feedback yielded numerous potential refinements and different perspectives, we selectively implemented only the most critical modifications that addressed some limitations without overly complicating the framework.

\begin{figure}[htb]
    \centering
    \includegraphics[width=1.0\textwidth]{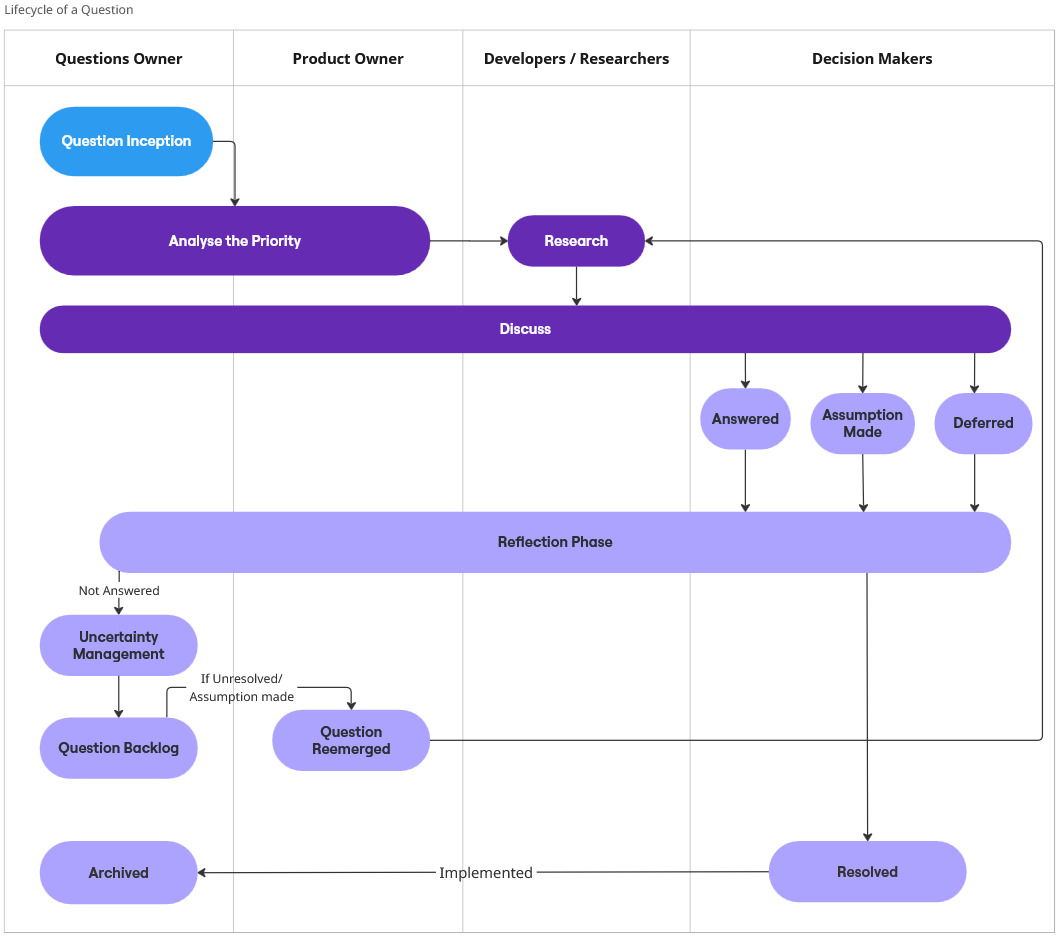} %
    \caption{Refined Lifecycle of a Question}
    \label{fig:question_lifecyclev2}
\end{figure}

The primary enhancement was the addition of an uncertainty management phase, which would follow the inference that a question couldn't be answered, and, in some cases, unanswered questions can be dealt with if their uncertainty is managed. Another instance of feedback received was the extension of the priority analysis, which was an activity to be held by the product owner and the questions owner. Finally, the re-emergence of a question would lead to another research phase before a broad discussion. This led to an improved lifecycle, shown in Figure~\ref{fig:question_lifecyclev2}.

\section{Future Work}


The future steps to fulfil the potential of this work include developing prototypes for an active assisted environment dedicated to managing natural questions. These prototypes will serve to validate the feasibility and effectiveness of the proposed lifecycle and requirements in real-world scenarios.

Additionally, future research will focus on assessing the impact of this approach by integrating it with industry-standard architecture frameworks and workflows and expanding its applicability to broader and broader software architecture and engineering contexts, ensuring that architectural question management becomes an essential practice in knowledge-driven development processes.

\section{Conclusion}


As knowledge continues to prove to be one of the most critical resources in the world, mismanaging it is a costly mistake. 
However, this persists as a challenge in software architecture, with a lack of tools and systems capable of systematically handling questions structurally and their dynamically intricate lifecycle. 
Natural questions shape systems and their inherent knowledge must be collected, organised and preserved with specialised tools. 

We identified critical gaps in existing knowledge solutions through a structured requirement-driven approach, including fragmented documentation practices, unspecialised question-tracking methods and poor integration with development workflows. To address these challenges we developed a structured lifecycle for managing architectural questions, attempting to capture the dynamic stages these questions go through from their inception to their resolve. This lifecycle provides a clear framework for how questions emerge, are discussed, refined, and ultimately resolved within software systems.

Building on this foundation, we identified a set of key requirements necessary for effectively supporting this lifecycle. These requirements were derived from an extensive literature review and an exploratory case study involving software architecture students and were later refined and validated through a survey with expert software architects.

By formalizing both the lifecycle and requirements for managing natural questions in software architecture, we provide a foundation for improving decision-making, preserving architectural knowledge and reducing the risks of knowledge loss. Our approach contributes to a more structured and reliable method for handling architectural questions, offering a practical framework that software architects can build upon and refine.

Future research will focus on further evaluating the effectiveness of the proposed lifecycle and requirements in industry settings, exploring prototyping and their integration into existing workflows, and ultimately assessing their long-term impact on architectural question preservation and decision traceability.

\subsubsection{Data Availability.}
Data from the workshop and survey will be made available in the ECSA
Zenodo community (https://zenodo.org/communities/ecsa).

\subsubsection{Disclosure of Interests.} The authors have no competing interests to declare
that are relevant to the content of this article.


\bibliographystyle{splncs04}
\bibliography{myRefs}

\end{document}